\documentclass[prb,aps,twocolumn,superscriptaddress,nofootinbib]{revtex4-2}

\usepackage{graphicx}
\usepackage{dcolumn}
\usepackage{bm}
\usepackage{bbm}
\usepackage{amsmath}
\usepackage{epsfig}
\usepackage{indentfirst}
\usepackage{psfrag}
\usepackage{subfigure}
\usepackage{amssymb}
\usepackage{color}
\usepackage{hyperref}
\hypersetup{colorlinks,allcolors=black}
\usepackage[T1]{fontenc}
\usepackage{comment}
\usepackage{upgreek}

\def\ii{{\rm i}}  \def\ee{{\rm e}}

\def\rb{{\bf r}}  \def\Rb{{\bf R}}

\def\kb{{\bf k}}  \def\Kb{{\bf K}}    
    
\def\Eb{{\bf E}}      
\def\pb{{\bf p}}   \def\Fb{{\bf F}} 
  
\begin{document}
\title{Laser-based aberration corrector}

\author{Zdeněk Nekula}
\email{zdenek.nekula@vutbr.cz}
\affiliation{Institute of Physical Engineering, Brno University of Technology, 616 69 Brno, Czech Republic}
\affiliation{Central European Institute of Technology, Brno University of Technology, 612 00 Brno, Czech Republic}
\author{Thomas Juffmann}
\affiliation{University of Vienna, Faculty of Physics, VCQ, A-1090 Vienna, Austria}
\affiliation{University of Vienna, Max Perutz Laboratories,
Department of Structural and Computational Biology, A-1030 Vienna, Austria}
\author{Andrea~Kone\v{c}n\'{a}}
\email{andrea.konecna@vutbr.cz}
\affiliation{Institute of Physical Engineering, Brno University of Technology, 616 69 Brno, Czech Republic}
\affiliation{Central European Institute of Technology, Brno University of Technology, 612 00 Brno, Czech Republic}

\begin{abstract}
Aberration correctors are essential elements for achieving atomic resolution in state-of-the-art electron microscopes. Conventional correctors are based on a series of multipolar electron lenses, but more versatile alternatives are intensively sought. Here we suggest spatially tailored intense laser pulses as one such alternative. Our simulations demonstrate that the free-space electron--photon interaction can be used to compensate for spherical and chromatic aberrations of subsequent electron lenses. We show a significant improvement in the simulated electron probe sizes and discuss the prospects of utilizing the tailored laser fields as a platform for novel electron optics in ultrafast electron microscope setups.  
\end{abstract}

\maketitle

\section{Introduction} 
The electron probes utilised in electron microscopes nowadays enable atom-by-atom structural and chemical sample analysis. However, the required sub-Å spatial resolution is not achieved routinely as the commonly used electro- and magneto-static electron lenses suffer from aberrations imposing deviations from ideal electron wavefronts. State-of-the-art (scanning) transmission microscopes ((S)TEMs) are limited by higher-order, mainly spherical aberrations, whose correction necessitates complicated correctors consisting of sets of magnetic multipole lenses \cite{Haider1998,Muller2006,Krivanek2008}. On the other hand, scanning electron microscopes (SEMs), low-energy electron microscopes or (S)TEMs operated at low acceleration voltages suffer from chromatic aberration, which can be either reduced by monochromation of the beam \cite{Freitag2005,PlotkinSwing2023}, compensated by a set of electric and magnetic multipoles \cite{Zach1995,Rose2005,Schmidt2010,Linck2016} or electrostatic mirrors \cite{Tromp2010,Tromp2013}. However, even electron microscopes equipped with correctors do not reach the theoretical $\sim$pm spatial resolution because the compensation is still imperfect \cite{Linck2016} and also limited by thermal noise \cite{Uhlemann2013}. Furthermore, the complexity, price, and lack of versatility of the commercially available correctors stimulate the search for alternative solutions.

Besides using reconstruction and computational techniques to improve the image resolution a posteriori \cite{kirkland2004indirect,jiang2018electron,nguyen2024}, the main alternatives for the multipole-based aberration correctors are electron phase plates (EPPs) \cite{malac2021phase}. One of the designs relies on a sculpted thin film with a tailored thickness profile imposing a local phase variation in the transverse electron wave function, compensating for the spherical aberration \cite{shiloh2018}. Another alternative design offering more versatility involves miniaturised einzel lenses that are operated as individual ``pixels'' imposing a local phase shift \cite{yu2023quantum}, which could correct for spherical aberration if a specific design is employed \cite{VegaIbanez2023}. 

Even more flexibility in modulating the electron beams is available when exploiting their interaction with tailored light. Very efficient electron-photon interaction can be reached when the electron passes through an evanescent optical field, which enables energy-momentum matching of free electrons and confined photons \cite{garciadeabajo2021_optical_excitations}. This mechanism is utilized in photon-induced near-field electron microscopy (PINEM), where the coherent control of fast electron wave function has been achieved through spatial dependence \cite{tsesses2023tunable}, intensity \cite{feist2015quantum}, temporal evolution \cite{vanacore2019ultrafast}, or quantum statistics \cite{dahan2021imprinting} of the optical near fields. In a configuration where a pre-shaped optical field produced, e.g., by a spatial light modulator (SLM), is reflected off a thin film opaque for light but transparent for electrons, the PINEM mechanism can serve as a versatile EPP. Theoretical predictions \cite{Konecna2020} showed that this setup could be used both in spherical aberration correction and in generating shaped electron beams, which was recently confirmed experimentally by demonstrating the preparation of Hermite-Gaussian beams \cite{Madan2022}.

\begin{figure*}
    \centering
    \includegraphics[width=0.9\textwidth]{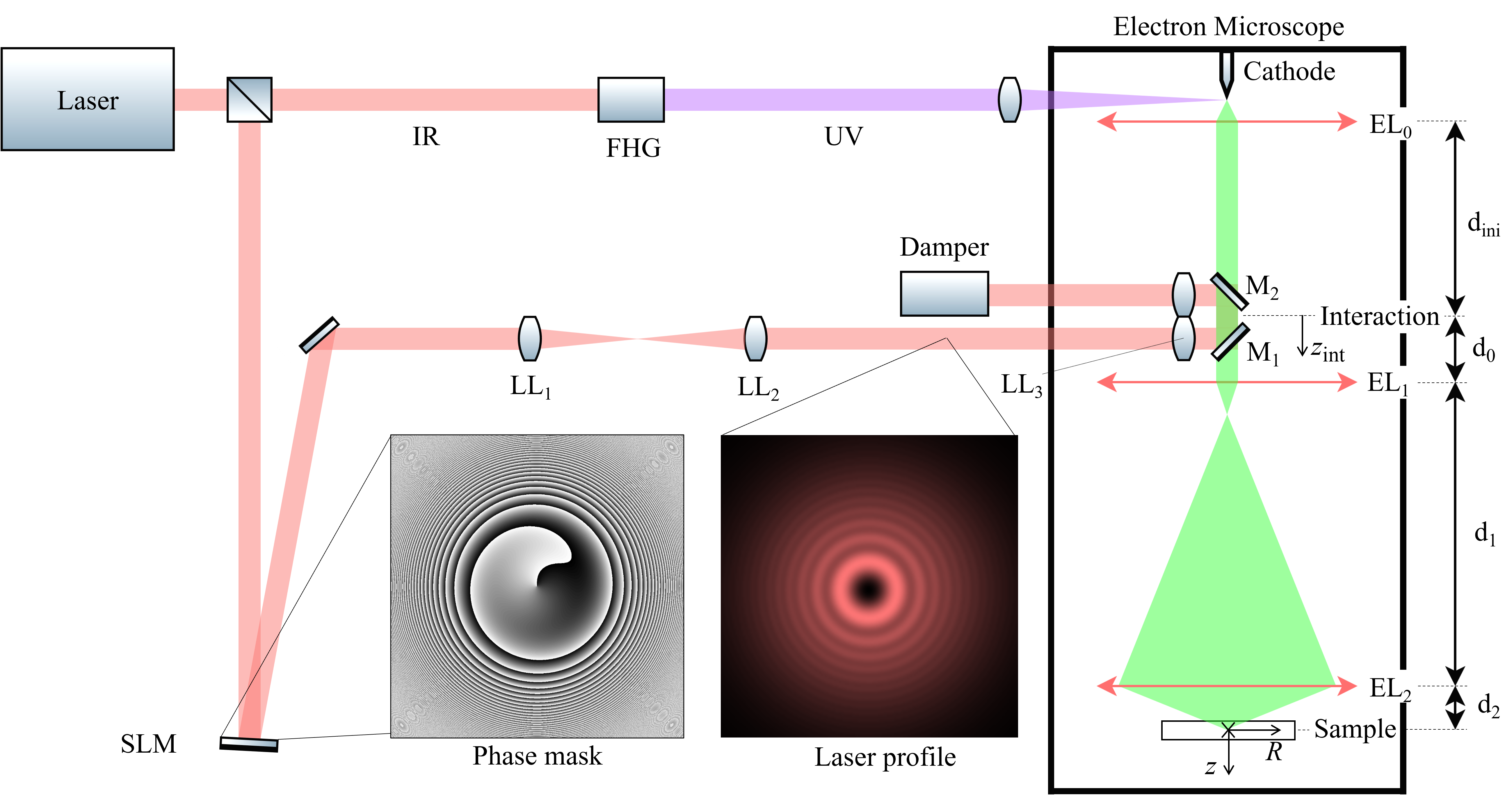}
    \caption{Scheme of a laser setup and an electron microscope. The femtosecond IR laser beam is divided into two branches. The first branch (top) undergoes the FHG, and the resulting UV beam is focused on a cathode to generate femtosecond electron pulses. The second laser branch is reflected from a spatial light modulator (SLM; see an example of a phase mask), and the tailored laser beam profile is focused into the interaction volume in an electron microscope by three lenses (LL$_1$, LL$_2$, LL$_3$). The laser axis coupling to the electron axis is carried out by a planar mirror (M$_1$). After passing the interaction volume, the laser beam is outcoupled from the microscope by the mirror M$_2$. Electron pulses enter and leave the interaction volume through a small drill in the mirrors. The interaction volume is centered around a conjugate plane to the electron objective lens (EL$_2$). We further denote the axes $z_\mathrm{int}$ and $z$ starting in the central plane of the interaction volume and in the sample plane, respectively. $R$ is the distance from the electron optical axis. In this scheme, dimensions are arbitrarily scaled for better visualization.}
    \label{fig:Fig1}
\end{figure*} 

 Unfortunately, the electrons' transmission through a thin film is likely to introduce unwanted beam distortion and decoherence, limiting the practical implementation in commercial setups. Such drawbacks are absent if electrons interact with light in free space. A well-known example is the diffraction of matter (electron) waves on standing light waves, the Kapitza-Dirac effect \cite{kapitza1933reflection,freimund2001observation} that has been applied as a Zernike EPP \cite{Schwartz2019,petrov2024crossed}. Consequently, an ``optical free-space electron modulator'' (OFEM) based on SLM-modulated optical pulses was suggested recently \cite{GarciadeAbajo2021}. In OFEM, the elastic non-linear free-space electron–light interaction is realized along the electron beam axis to achieve wave function modulation in the plane transverse to the optical axis. While this process requires high laser intensities to enable the underlying two-photon process, proof-of-concept experiments demonstrated its feasibility in an ultrafast SEM setup, where intense femtosecond laser pulses were capable of generating variously shaped electron beam profiles or even acting as convergent and divergent lenses \cite{Mihaila2022}. The use of specifically shaped (e.g., Laguerre-Gauss or Bessel-like optical beams) for electron lensing and spherical aberration correction was also discussed theoretically \cite{Uesugi2021,GarciadeAbajo2021,ChiritaMihaila:25}. 

In this Letter, we exploit the full potential of the free-space electron-photon interaction to introduce a conceptually new chromatic aberration corrector -- on par with Ref.~\cite{ChiritaMihaila2025_chromatic} -- and further study the capabilities of the spherical aberration corrector beyond the above-mentioned works. We simulate a virtual ultrafast electron microscope accompanied by an optical setup with realistic parameters and theoretically demonstrate that the light-based aberration correctors could complement or even outperform the standard aberration correctors in terms of the microscope's resolution improvement, physical dimensions, and versatility.

\section{Virtual electron microscope}
The scheme of the virtual electron microscope with a light-based aberration corrector is shown in Fig.~\ref{fig:Fig1}. We adopt the strategy of Ref.~\cite{Mihaila2022} and assume a laser source capable of producing very intense quasi-monochromatic infrared (IR) femtosecond pulses with central wavelength $\lambda_\mathrm{L}=1035$~nm. We approximate the corresponding electric field propagating along the optical axis in the $\hat{z}$ direction as
\begin{align}
    \Eb(\rb,t)&\approx \hat{x}\,\mathrm{Re}\lbrace E_\perp(\rb)\rbrace\,g(t\pm z/c)\, \mathrm{sin}(\omega_\mathrm{L} t+\phi_0),
    \label{eq:laser_field}
\end{align}
where $\rb=(x,y,z)$ is the position vector, $E_\perp$ is a complex field amplitude, $g$ characterizes the Gaussian pulse envelope, $\omega_\mathrm{L}$ is the angular frequency of the field, and $\phi_0$ is a constant. 

The initial laser pulses are split into two branches. The first branch is used for the generation of electron pulses via the photoemission process and the second branch is responsible for the aberration correction. The photoemission process requires sufficient energy of individual photons, which can be achieved through fourth-harmonic generation (FHG). The resulting ultraviolet (UV) photons are focused on a cathode. In our virtual setup, we assume that electrons are emitted at random times within the laser pulse duration and that they have a finite initial energy spread uncorrelated with the emission time. We do not simulate the photoemission nor the acceleration process. Instead, we assume that electrons in a certain $xy$ plane after the acceleration region (just after the lens EL$_0$), are homogeneously spread over a disk with a diameter of 10~$\upmu$m. Due to the uncertainty in the exact emission time, the electrons also have a Gaussian temporal spread with FWHM $\Delta \tau=200$~fs, which is further increased during their propagation over the distance $d_\mathrm{ini}$ towards the interaction volume. 

The second laser branch is responsible for generating a suitable laser intensity profile in the electron–light interaction region. It contains a set of optical lenses and an SLM, whose action is described in terms of a phase imprinted in the incident electric field, which initially has a Gaussian intensity profile in the $xy$ plane. Propagation of light between the individual elements is performed using the angular spectrum method (ASM; see Appendix \ref{App:laser_propagation}). The Fourier plane of the SLM is imaged into the center of the interaction volume by three lenses (LL$_1$, LL$_2$, LL$_3$). Two mirrors inside the electron-microscope column are responsible for aligning the optical axes of the second light branch and the electron microscope. However, small holes have to be drilled in the mirrors' centers to allow the narrow electron beam to pass through the mirrors while the broader laser beam gets reflected on the surrounding mirror surface. 

Propagation of electrons within the virtual electron microscope relies on solving relativistically-corrected equations of motion for point-like charged particles. The electrons move along straight trajectories representing rays until they encounter one of the electron lenses or the interaction volume where we focus the laser pulses (see Appendix \ref{App:electron_propagation} for details). The electromagnetic field in Eq.~\eqref{eq:laser_field} allows for the perturbative treatment of the equation of motion, where we split fast and slow variations in electron position and momentum (associated with fast field oscillations at the optical frequency and slower evolution of its Gaussian envelope). If the electron moves within the field for many field cycles, the overall action of the electromagnetic field can be time-averaged over the optical field period to yield an effective ponderomotive force acting on the electrons. The averaged electron momentum evolution is obtained by solving \cite{Quesnel1998}
\begin{align}
\mathrm{d}\bar{\mathbf{p}}(\rb,t)/\mathrm{d}t=-\frac{e^2}{4 \gamma m_\mathrm{e} \omega^2}\nabla \left\lvert \Eb_\perp(\rb)g(t\pm z/c)\right\rvert^2,
\label{Eq:trajectory_eq}
\end{align}
where $m_\mathrm{e}$ is the electron rest mass, $e$ the elementary charge, and  $\gamma=\sqrt{1/(1-(\dot{\rb}/c)^2)}$ is the Lorentz factor with $c$ the speed of light in vacuum. The $\pm$ sign is chosen based on the direction of propagation of the laser with respect to electrons moving in the positive $\hat{z}$ direction ($+/-$ for counter-/co-propagating beams). We note that Eq.~\eqref{Eq:trajectory_eq} is approximate and holds only for small field magnitudes (such that $e \lvert \Eb_\perp\rvert/(m_\mathrm{e} c \omega)\ll 1$). We can also easily see that if an electron travels through the optical field, it is steered in the opposite direction with respect to the increasing gradient of the laser intensity. 

As soon as the laser intensity in the interaction volume is known, we solve Eq.~\eqref{Eq:trajectory_eq} numerically by tracing the electrons over sufficiently small steps in time using the relativistically corrected leapfrog method \cite{Hairer2023}. The action of conventional electron lenses (EL$_1$, EL$_2$) is described in terms of a sudden imprinted phase and an associated momentum kick at the corresponding position along the $z$ axis. We assume that the lens EL$_1$ is used to image the plane centered in the electron-light interaction volume onto the plane of the objective lens (EL$_2$). Therefore, all changes of the electron phase induced by the laser in the interaction volume are as if they happen directly in the objective lens. 

In the following, we show that the momentum acquired by the electrons due to the ponderomotive force can be utilized to compensate for the most pronounced aberrations of a conventional objective lens: the spherical and chromatic aberrations. In both cases, we typically express the distortion caused by the aberrations within a phase of the electron wavefront. A (thin) lens with a focal distance $f$ and the spherical aberration coefficient $C_\mathrm{S}$ (which is always positive for static, round and energy-preserving electron lenses \cite{Scherzer1936}) imposes the phase variation
\begin{align}
    \varphi(R)=
    \frac{qR^{2}}{2f} + 
    \frac{C_\mathrm{S} q R^{4}}{4 f^{4}},
    \label{eq:lens_focusing_spherical}
\end{align}
where the first term stands for the focusing action of the lens, while the second term emerges due to the spherical aberration. $R=\sqrt{x^2+y^2}$ denotes the distance from the optical axis and $q$ is the electron wave vector. Furthermore, the chromatic aberration causes the focal length to depend on the acceleration voltage  
\begin{align}
    f(V) = f_0 + C_\mathrm{C} \frac{V - V_0}{V_0},
    \label{eq:chromatic}
\end{align}
where $f_0$ is the nominal focal distance, $C_\mathrm{C}$ is the chromatic aberration coefficient, and $V_0$ is the nominal acceleration voltage, while $V$ is the actual acceleration voltage of individual electrons.

\begin{figure*}
    \centering
    \includegraphics[width=0.88\textwidth]{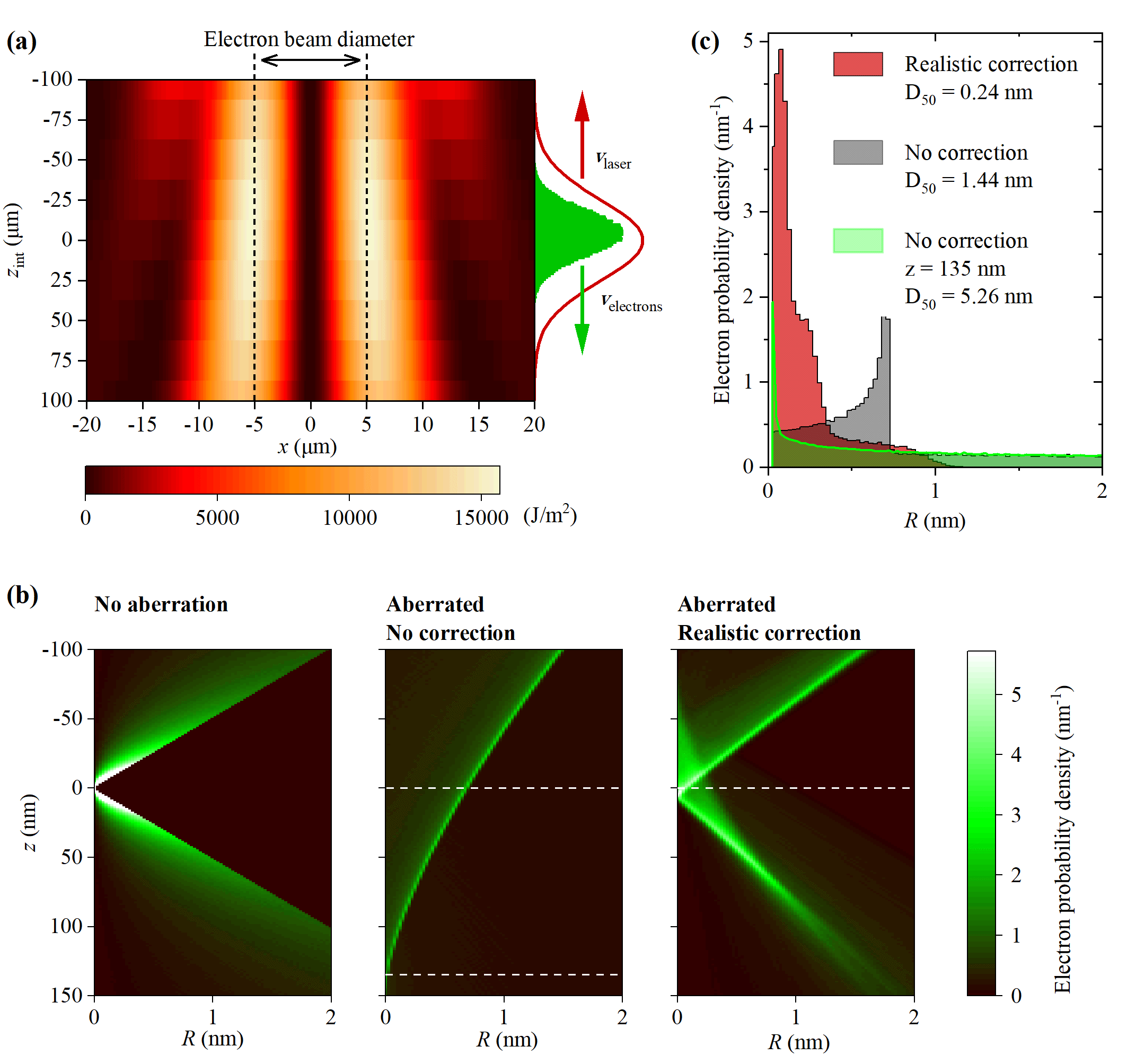}
    \caption{Spherical aberration correction in a virtual 200-kV electron microscope. 
    \textbf{(a)} Cross-sections of laser planar energy density (see text) in the interaction volume. The green histogram visualizes the distribution of electrons in a pulse while the red curve shows laser pulse intensity distribution along $z_\mathrm{int}$ at $t=0$. The overall laser pulse energy is 4.05~$\upmu$J.
    \textbf{(b)} Histograms of the electron distribution in space close to the sample plane $z=0$ for a non-aberrated microscope (left), the microscope with an objective with spherical and chromatic aberrations (middle), and the microscope with the aberrated objective, but equipped with a laser-based corrector (right).
    \textbf{(c)} Histograms of electron distributions in the aberrated (gray) and corrected (red) setups at the sample plane $z=0$. Both cases were optimized to get the smallest spot diameter (D$_{50}$). 
    D$_{50}$ is decreased by a factor of six using the laser-based aberration corrector. The green histogram shows an uncorrected electron distribution in plane $z = 135$~nm. 
    }
    \label{fig:Fig2}
\end{figure*}

The propagation of electrons after their transmission through the lens can be expressed in terms of the spatial evolution of wavefronts but we can also translate the imprinted phase to trajectories while neglecting diffraction. In free space, the trajectories are perpendicular to wavefront surfaces and the velocity vector components $\mathbf{v}=(v_x,v_y,v_z)$ just after the transmission through the lens can be written as
\begin{align}
    \mathbf{v}&\approx\mathbf{v}_0-\frac{v_{z,0}}{q}\,\left( \frac{\partial\varphi}{\partial x} ,\frac{\partial\varphi}{\partial y} ,0\right),
    \label{Eq:velocity_phase_lens}
\end{align}
where $\mathbf{v}_0=(v_{x,0},v_{y,0},v_{z,0})$ is the electron velocity just before the lens, and where we assumed the validity of the paraxial approximation, in which the electron velocity component $v_z\gg v_x, v_y$ and where the phase variation due to the presence of the lens only introduces a small perturbation. In the calculations, we further multiply Eq.~\eqref{Eq:velocity_phase_lens} by a constant ensuring that the velocity magnitude does not change. 

\section{Compensation of spherical and chromatic aberrations} 

\begin{figure*}
    \centering
    \includegraphics[width=0.88\textwidth]{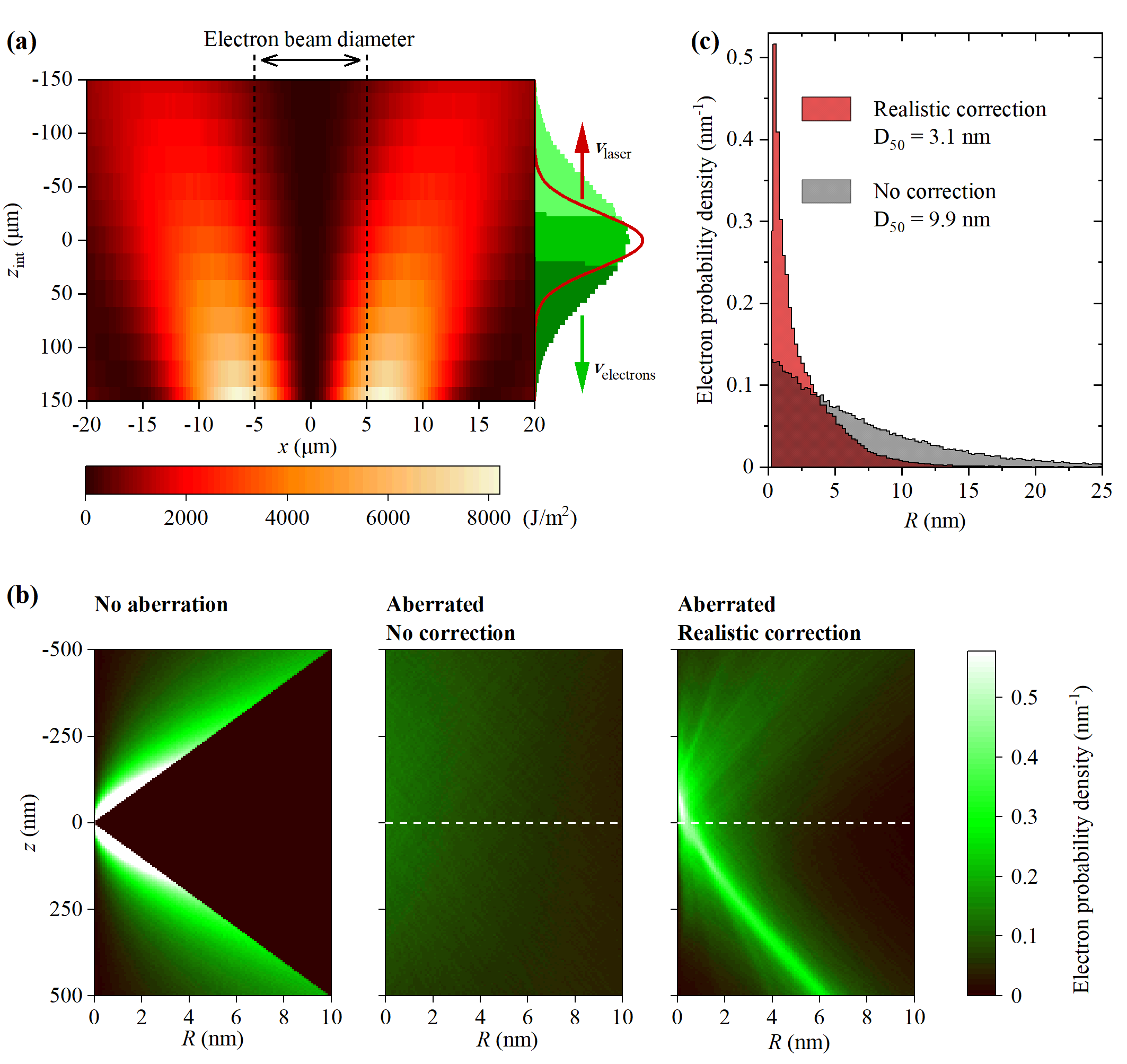}
    \caption{Chromatic aberration correction in an electron microscope operating at low acceleration voltage (500~V). See Table~\ref{tab:Tab1} for a complete list of setup parameters. \textbf{(a)} Cross-sections of laser planar energy density in the interaction volume. Laser pulse energy is 3.9~$\upmu$J. The histogram shows the energy distribution within the electron pulse in the interaction region: dark/bright green corresponds to the most/least energetic electrons. The red curve shows laser pulse intensity distribution along $z_\mathrm{int}$. 
    \textbf{(b)} Histograms of the electron distribution around the sample plane $z=0$ evaluated in analogous manner to Fig.~\ref{fig:Fig2}(b). 
    \textbf{(c)} Histogram of electron distribution at $z=0$ in the aberrated (gray) and corrected (red) setup. 
    }
    \label{fig:Fig3}
\end{figure*}

We first simulate the possibility of using the ponderomotive force to compensate for aberrations in an ultrafast transmission electron microscope operating at a higher acceleration voltage (200~kV in this example). The resolution in such a setup is typically limited by the spherical aberration of an objective lens (EL$_2$). In our demonstration, we consider a convergence semiangle of 19.9 mrad, and $C_\mathrm{S}=1$~mm based on Ref.~\cite{Krivanek2008}. The properties of the electron source are set according to Ref.~\cite{Feist2017}. A complete list of parameters characterizing this setup can be found in Table \ref{tab:Tab1} in Appendix \ref{App1:Table}. 

Using the analytically calculated laser profile (see Eq.~\eqref{eq:analytical_spherical} in Appendix \ref{App:analytical_solutions}), we find that the radial profile of the laser beam intensity corresponding to the Laguerre-Gauss LG$_{01}$ mode is naturally very close to the intensity distribution needed to compensate for the spherical aberration and the defocus. Moreover, this doughnut-shaped mode undergoes sufficiently low losses upon reflection on the drilled mirror. The optimization process to fine-tune the beam parameters is described in Appendix \ref{App:optimization}. The leading and tailing fractions of the electron pulse will interact with the incoming laser pulse in slightly different planes $z_\mathrm{int}$. This leads to altered integrated laser powers, which we call planar energy density and which is shown in Fig.~\ref{fig:Fig2}(a). We then utilize electron tracing to calculate the electron distribution in different planes close to the sample plane in Fig.~\ref{fig:Fig2}(b). We note that the distribution is already integrated over the azimuthal direction. The left histogram is obtained for an ideal system without aberrations, where the resulting beam is focused to a point (the diffraction limit would be 0.077 nm in this case, but it is not considered within the tracing approach). If we introduce the spherical and chromatic aberration to the objective lens, the electron beam is significantly distorted and broadened as shown in the middle histogram. On the other hand, the interaction with the laser field (prepared with the SLM phase plotted in Fig.~\ref{fig:Fig1}) can partially compensate for the unwanted distortions of the electron distribution as demonstrated in the right histogram. To facilitate the comparison, we plot the probe profiles in the sample plane (at $z=0$) for the aberrated (gray) and laser-corrected (red) case in Fig.~\ref{fig:Fig2}(c). By comparing the probe diameter containing 50\% of all electrons (D$_{50}$), which is the parameter we optimized for, we find that the correction yields an improvement of the probe size by a factor of six in diameter. By checking the evolution of the aberrated case along the $z$ axis in (b), it might seem that the probe size improves in a defocused plane. However, despite the narrow central peak, most of the electrons are spread over an extended area, causing unwanted beam ``tails'' and a significantly increased value of D$_{50}$ as shown in the green histogram in (c).

We now consider a low-voltage electron microscope, where the resolution is mostly restricted by chromatic aberration. In simulations presented in Fig.~\ref{fig:Fig3}, we assume 500-eV electrons with a Gaussian energy distribution of FWHM 0.6~eV. This polychromatic electron pulse spreads along its trajectory from the gun to the interaction volume (along the distance $d_\mathrm{ini}=200$~mm). We take advantage of the fact that the more energetic electrons travel faster and form the leading edge of the pulse, while the less energetic electrons are slower and create a trailing pulse edge. In this way, electrons in the pulse sort themselves according to their energy along the optical axis. We illustrate this concept in the sketch in Fig.~\ref{fig:Fig3}(a), where the faster electrons are represented by dark green, while the slower ones by bright green. This velocity-related separation enables individual treatment of electrons according to their energy in order to compensate for the chromatic aberration of the objective lens. The corresponding laser planar energy density (and also intensity) profiles plotted in Fig.~\ref{fig:Fig3}(a) have parabolic shapes in their center and act as a series of electron lenses with varying focal distances. The effective focal length of those laser-based lenses is getting shorter along the optical axis. Therefore, faster electrons at the leading edge are focused more, which compensates for the longer effective focal length of the chromatically-aberrated objective. In contrast, slower electrons at the trailing edge are focused less, compensating for the shorter effective focal length of the objective. 

After tracing the electrons through the laser intensity and the rest of the microscope setup using the same procedure as in the case of 200-keV microscope, we can again compare the electron histograms in the volume around the sample plane in Fig.~\ref{fig:Fig3}(b). Comparison of the spot size in the ideal non-aberrated setup (left histogram) with the setup with an aberrated objective (middle histogram) clearly demonstrates that the aberrations, and dominantly the chromatic aberration, significantly broaden the ideal electron spot size in the radial direction but also along the optical axis. On the other hand, the laser-corrected probe (right histogram) gets again more localized. If we evaluate the aberrated and corrected probe diameters D$_{50}$ in the sample plane [see Fig.~\ref{fig:Fig3}(c)], we get an improvement by a factor of three. 

We note that the performance of both simulated setups will also depend on the precision of the synchronization between the electron and laser pulses and on laser energy fluctuations. We find that the probe diameter D$_{50}$ only increases by 10 \% in the 200-kV setup (Example 1) and by 0.4 \% in the low-voltage setup (Example 2) if we consider the synchronization error in the arrival of the laser and electron pulses of $\pm 100$ fs. The laser energy variation of $\pm 1$ \% of the nominal energy leads to a 4 \% probe size increase in Example 1 and 0.4 \% in Example 2. Such stability can easily be reached with current technology, which confirms the robustness of the laser-based aberration compensation.

\section{Conclusions}
In conclusion, we have introduced a novel approach for simulating electron microscopes equipped with a free-space parallel laser-electron interaction module, demonstrating its application in two representative setups: one operating at high and the other at low acceleration voltages. Our simulations incorporate realistic capabilities and limitations of laser and electron optics, and indicate that the laser-based correction can significantly improve the electron microscope performance in both cases. Specifically, for the high-voltage transmission electron microscope, the correction effectively reduces spherical aberration and the impacted probe size by a factor of six. In the low-voltage setup, our novel approach leverages the natural dispersion of electrons within the interaction volume to eliminate the influence of the dominant chromatic aberration, and reduces the probe size by a factor of three.

The predicted improvements are comparable to those achieved with existing commercial solutions, but further optimization of the laser profiles remains possible. The presented results are not yet at the theoretical optimum nor laser-wavelength diffraction limited, suggesting that iterative optimization algorithms, such as gradient descent methods with feedback, could offer substantial improvements \cite{Schroff2023}. With more refined adjustments, it may be possible to correct higher-order aberrations and accommodate larger convergence angles, expanding the corrective capabilities of this technique. 

Although the applicability of the laser-based aberration correction is currently limited due to high laser pulse energies achievable only with femtosecond lasers, it still represents a promising and compact solution for improving the imaging performance of ultrafast electron microscopes. The strategic placement of the laser interaction volume in the conjugate plane of the objective considered in our virtual setup not only facilitates aberration correction but also holds promise for additional on-demand electron beam shaping \cite{GarciadeAbajo2021}.  
Future work aimed at optimizing laser profiles and extending the technique to higher-order aberrations will further elevate the capabilities of this method, contributing to the next generation of versatile electron microscopes.

\begin{acknowledgments} 
This work has been supported by the Czech Science Foundation GACR under the Junior Star grant No. 23-05119M). TJ acknowledges funding from the European Union’s Horizon 2020 research and innovation program under Grant Agreement No. 101017902.
\end{acknowledgments}

\bibliographystyle{apsrev4-2}
%

\clearpage
\onecolumngrid
\appendix

\setcounter{equation}{0}
\renewcommand{\theequation}{A\arabic{equation}}

\section{Parameters of simulated examples} 
\label{App1:Table}
In Table \ref{tab:Tab1} we list all parameters used in simulations. 

\begin{table*}[h]
\caption{
Parameters used for the two model examples introduced in the main text. Example 1 (Fig.~\ref{fig:Fig2}): the spherical aberration correction is demonstrated on a 200 kV STEM with objective parameters from Ref.~\cite{Krivanek2008} and the electron source corresponding to Ref.~\cite{Feist2017}. Example 2 (Fig.~\ref{fig:Fig3}): the chromatic aberration correction assumes the same electron-microscope parameters except for the acceleration voltage, which was decreased down to 500 V. For labeling of different optical and electron-optical elements, see Fig.~\ref{fig:Fig1}. Electron particle tracing starts just after plane EL$_0$. Here, the electron beam is defined by initial conditions: it is collimated, pulsed, and electron energy spectra have a Gaussian distribution, but electrons with different energies are randomly positioned within the pulse. The defocus is the distance between the sample plane ($z$ = 0) and a plane where the paraxial electrons are focused. The effect of electron objective EL$_2$ defocus can be clearly seen in Fig.~\ref{fig:Fig2}(c) (Aberrated, No correction).
}
\begin{tabular}{lll}
\hline\hline
& \textbf{Example 1 }           & \textbf{Example 2}     \\ \hline\hline
\textbf{Electron microscope and beam parameters} & &\\ Nominal acceleration voltage & 200 kV & 500 V \\
Electron energy spread (Gaussian, FWHM) & 0.6 V & 0.6 V \\
Initial electron pulse distribution in time (Gaussian, FWHM) & 200 fs & 200 fs \\ 
Electron beam width in the interaction volume & 10 $\upmu$m & 10 $\upmu$m \\
Initial electron pulse distribution in space & top-hat, 10 $\upmu$m width   & top-hat, 10 $\upmu$m width \\
Electron convergence semiangle & 19.9 mrad & 19.9 mrad \\
Demagnification (beam diameter ratio on EL$_2$ and EL$_1$) & 6  & 6 \\
EL$_1$ focal distance  $f_1$ & 37.5 mm & 37.5 mm \\
EL$_2$ focal distance  $f_2$ & 1.500381 mm & 1.5015 mm \\
EL$_2$ defocus for uncorrected & 0.148 $\upmu$m & 0.0 $\upmu$m \\
EL$_2$ defocus for   laser-corrected & 1.3798 $\upmu$m & 1.5 $\upmu$m \\
EL$_1$ $C_\mathrm{S}$ & 0 mm & 0 mm \\
EL$_2$ $C_\mathrm{S}$ & 1 mm  & 1 mm \\
EL$_1$ $C_\mathrm{C}$ & 0 mm & 0 mm \\
EL$_2$ $C_\mathrm{C}$ & 1.1 mm & 1.1 mm \\
$d_\mathrm{ini}$ (gun to the interaction volume distance) & 200 mm & 200 mm \\
$d_0$ (interaction volume to EL$_1$ distance) & 43.75 mm & 43.75 mm \\
$d_1$ (EL$_1$ to EL$_2$ distance) & 262.5 mm & 262.5 mm \\
$d_2$ (EL$_2$ to sample plane distance) & 1.510067 mm &  1.510067 mm \\
Diffraction limit of the probe size & 0.077 nm & 1.7 nm\\ \hline

\textbf{Laser setup and beam parameters} & &\\ 
Laser pulse propagation direction & opposite to electrons & opposite to   electrons \\
Laser pulse duration (Gaussian, FWHM) & 200 fs & 200 fs \\
Laser wavelength & 1035 nm & 1035 nm \\
Laser pulse energy (in numerical simulations) & 4.05 $\upmu$J & 3.9 $\upmu$J \\
LL$_1$ focal distance & 200 mm & 200 mm \\
LL$_2$ focal distance & 370 mm & 500 mm \\
LL$_3$ focal distance & 40 mm & 40 mm\\
Mirror drill diameter & 200 $\upmu$m & 200 $\upmu$m \\
Distance SLM to LL$_1$ & 200 mm & 200 mm \\
Distance LL$_1$ to LL$_2$ & 570 mm & 700 mm \\
Distance LL$_2$ to LL$_3$ & 150 mm & 150 mm \\
LL$_1$, LL$_2$, LL$_3$ diameter & 25 mm & 25 mm \\
SLM size & 9.6 × 9.6 mm$^2$ & $9.6 \times 9.6$ mm$^2$ \\
SLM pixels & $1200 \times 1200$ & $1200 \times 1200$ \\ \hline

\textbf{Simulation parameters} && \\
Number of traced electrons & 100000 & 100000 \\
Time step & 10 fs & 50 fs \\
Number of pixels in laser wave propagation space & $16000 \times 16000$ & $16000 \times 16000$ \\
Space size  in laser wave propagation & $25 \times 25$ mm$^2$ & $25 \times 25$ mm$^2$ \\
Scaling factor $\sigma$ & 0.1  & 0.2\\ \hline\hline       
\end{tabular}
\label{tab:Tab1}
\end{table*}

\section{Laser wave propagation} 
\label{App:laser_propagation}
\subsection{Angular spectrum method and scaling method}
Calculation of laser intensity distribution in the interaction volume requires propagation of the complex laser amplitude $E_\perp$ through the laser optics, that is, from the SLM through lenses and the mirror with a drilled hole to the interaction volume along the path sketched in Fig.~\ref{fig:Fig1}. To propagate the complex wave between the individual elements, we use the angular spectrum method (ASM). The ASM relies on the decomposition of $E_{\perp,0}$ in an initial plane $z_0$ into a spectrum of plane waves by Fourier transformation
\begin{align}
    E_\perp(\Kb,z_0)=\int\mathrm{d}^2\Rb\,E_{\perp,0}(\Rb,z_0)\,\ee^{-\ii\Kb\cdot\Rb},
\end{align}
where $\Kb=(k_x,k_y)$ are transverse components of the light wave vector $\kb$. Each plane-wave spectral component is then propagated along a desired distance along the optical axis in the $z$ direction, which means it acquires phase $k_z z$ with $k_z=\sqrt{k^2-K^2}$, i.e., $E_\perp(\Kb,z)=\ee^{\ii k_z z}E_{\perp,0}(\Kb,z_0)$. Then, the resulting propagated wave in a plane $z$ can be expressed as a function of real-space coordinates by performing an inverse Fourier transform
\begin{align}
    E_\perp(\Rb,z)=\int\int  \frac{\mathrm{d}^2\Kb\mathrm{d}^2\Rb'}{(2\pi)^2}\ee^{\ii(\Kb(\Rb-\Rb')+k_z z)}E_{\perp,0}(\Rb',z_0).
\end{align}

The ASM can not be easily replaced by ray tracing since ray tracing cannot capture interference effects, possibly affecting intensity patterns in the interaction volume. In practice, the Fourier transforms are replaced by discrete fast Fourier transform (FFT) algorithms. The ASM thus samples each plane perpendicular to the optical axis by a square grid of pixels. The size of the grid has to be at least equal to or larger than the largest aperture to prevent artefacts due to the boundaries. In our case, we consider the spatial grid of 25 × 25 mm$^2$ (as our setup features apertures with 25 mm diameter). However, the laser is focused to a diameter of approximately 10 $\upmu$m in the interaction volume. If the laser intensity in the interaction volume should have sufficient sampling of at least 0.2 $\upmu$m, it would require $125000\times 125000$ pixels. This means the calculation of 2D FFTs on 125 GB matrices. Although this task is solvable for computers with a large operation memory, it is a limiting aspect of these simulations. Moreover, this calculation takes a long time, depending on the laser setup's complexity and the number of planes towards which the field is propagated. Such a simulation usually takes multiple hours (of serial computing), which is a problem for iterative calculations and optimization. For this reason, we found an approximate solution called a scaling method. 

The scaling method is inspired by the natural properties of a spherical wave. Since the focused laser beam is similar to the spherical wave, we can treat it as a spherical wave with small perturbations. In the laser setup, the final objective lens (LL$_3$) focuses the laser beam into the interaction volume, where a high spatial resolution is required. The distance between the objective lens and the crossover is a working distance (WD). In the scaling method, we substitute the original objective lens with the working distance WD with a lens with a longer working distance $\overline{\mathrm{WD}}$. The ratio of the original and substituted working distance is the scaling factor $\sigma$
\begin{align}
    \sigma = \mathrm{WD} \,/\, \overline{\mathrm{WD}}.
\end{align}
The substituted objective lens will create a similar laser spot at its real working distance but with a larger size as described by
\begin{align}
    E_\perp (x,y,z)  \approx \frac{1}{ \sigma } \overline{ E_\perp }  \left( \frac{x}{ \sigma },\frac{y}{ \sigma },\frac{z}{ \sigma^{2} } \right),
\end{align}
where $E_\perp$ ($\overline{E_\perp}$) is the complex laser amplitude in the vicinity of the objective lens' (substituted lens') working plane at $z = 0$. In our case, the working distance is equal to the focal length. The focal length and the working distance are equal since the incoming beam is collimated. Therefore, $\sigma = f \,/\, \overline{f}$ is valid as well.

\begin{figure}
    \centering
    \includegraphics[width=0.5\linewidth]{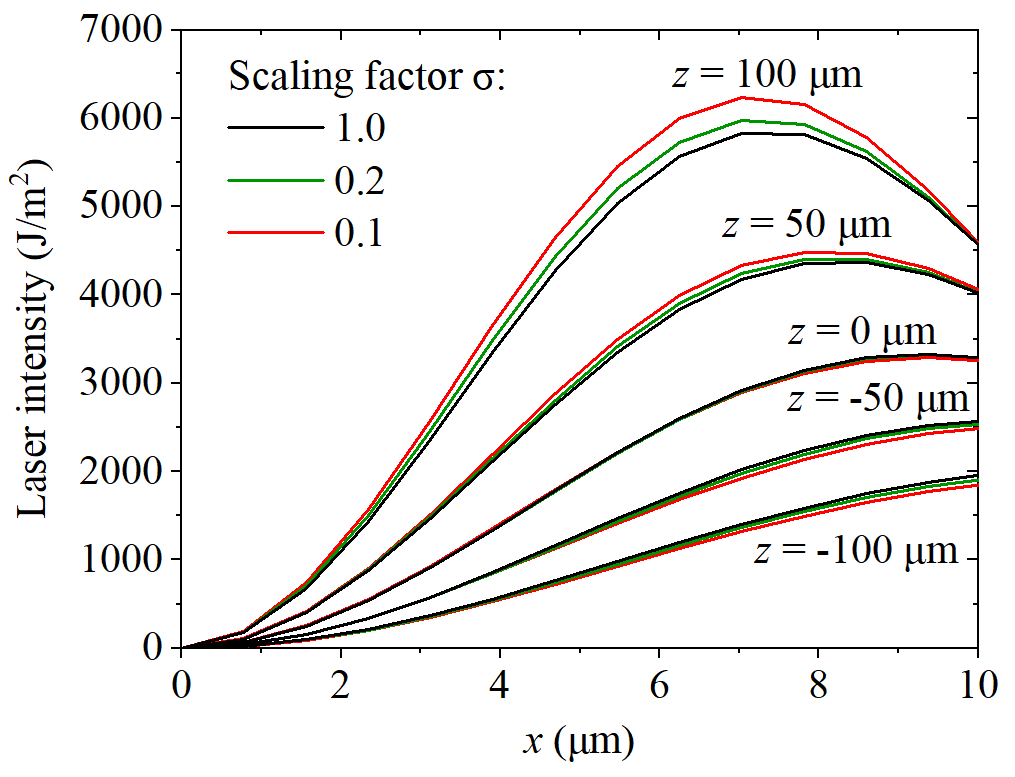}
    \caption{Cross-sections of a focused laser beam into the interaction volume in various planes along the optical axis calculated by propagating the complex laser wave through laser optics with different scaling factors. We can see that the more $\sigma$ differs from 1 (no scaling), the bigger the error in the resulting intensity cross-section. The error in intensity creates a small error in the calculation of the corrected electron spot size D$_{50}$. In this particular case, corresponding to the low-voltage electron microscope (see also Fig.~\ref{fig:Fig3}) (Example 2), the simulated electron spot size errors induced by the scaling method are 0.2 \% for $\sigma$ = 0.2 and 0.4 \% for $\sigma$ = 0.1.}
    \label{fig:Fig_scaling}
\end{figure}

The advantage of the scaling method is a simplification of ASM numerical calculation. For example, choosing a scaling factor of 0.1 will decrease the number of pixels in the simulated space by 100 times. This makes the ASM simulations much faster and far less demanding on operating memory. In our examples, we used a scaling factor equal to 0.1 and 0.2. The influence of the scaling method on the accuracy of the resulting data is shown in Fig.~\ref{fig:Fig_scaling}.

We also assume that the laser pulse envelope propagates in time as
\begin{align}
    g^2(t) = \frac{1}{s \sqrt{2\pi}} \mathrm{exp} \left( -\frac{t^2}{2s^2} \right),
\end{align}
where the standard deviation $s$ can be calculated from the full width at half maximum (FWHM) of the laser pulse as $s=\mathrm{FWHM}/(2\sqrt{2\mathrm{ln}2})\approx \mathrm{FWHM}/2.3548$. For visualization of the numerically calculated laser fields, we define the ``laser planar energy density'', which is the quantity plotted in Fig.~\ref{fig:Fig2}(a) and Fig.~\ref{fig:Fig3}(a), as
\begin{align}
    \eta(\rb) =\frac{c\,\varepsilon_0}{2}\lvert E_\perp(\rb)\rvert^2,
\end{align}
where $\varepsilon_0$ is the vacuum permittivity, and $c$ is the speed of light in vacuum.

\subsection{Laser setup elements} 
The first simulated optical element sketched in Fig.~\ref{fig:Fig1} is the SLM. Each SLM pixel induces a phase shift in the range  $\langle 0;2 \pi )$. In our simulation, the SLM is an ideal device with 100 \% modulated reflected light. The laser beam impinges and reflects perpendicular to the SLM screen. 

The lenses (LL$_1$, LL$_2$, LL$_3$) placed after the SLM are ideal thin lenses described by the phase modulation
\begin{align}
    \varphi_\mathrm{light}(R)=\frac{kR^{2}}{2f_\mathrm{LL}},
    \label{eq:light_lens_phase}
\end{align}
where $R=\sqrt{x^2+y^2}$ is the distance from the optical axis, and $f_\mathrm{LL}$ is the focal distance of the light lens. 

\begin{figure}
    \centering
    \includegraphics[width=0.5\linewidth]{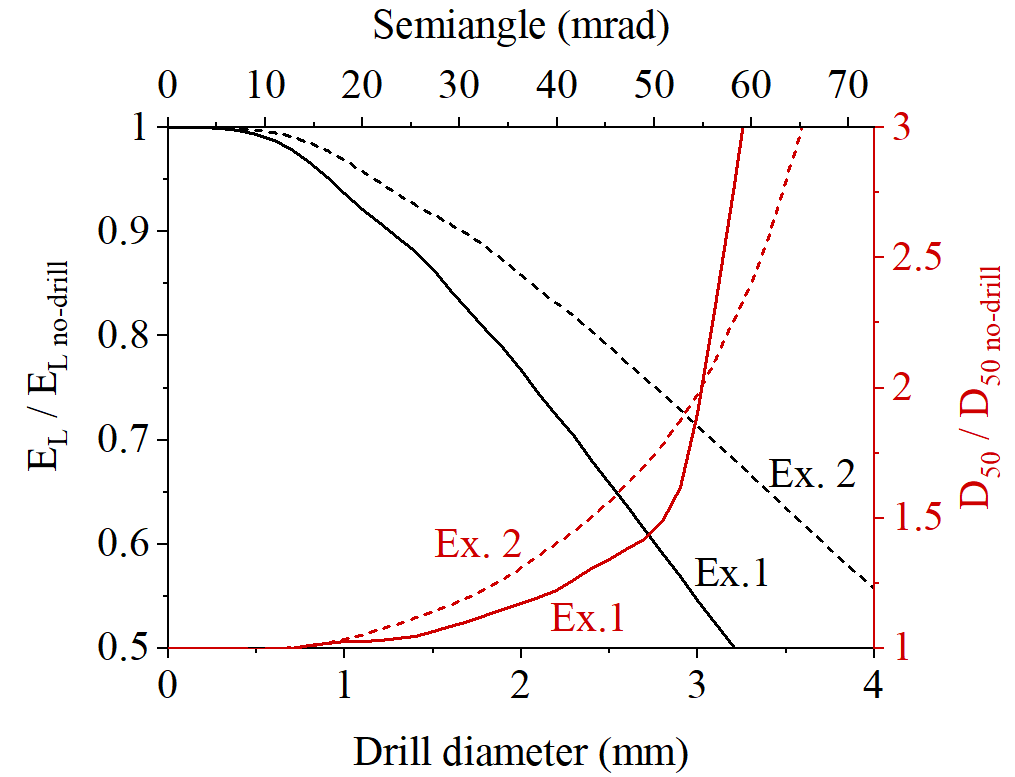}
    \caption{Influence of a mirror drill size on the laser-based aberration corrector performance. In our calculations, the mirror center is positioned 27.5 mm from the center of the interaction volume (from which we can calculate the semiangle obscured by the drill). If the drill is small enough, it has a negligible influence since the intensity in the center of the doughnut-shaped beams employed in our simulations is close to zero. The black curves show the ratio of reflected laser energy of the laser beam with respect to the total intensity in our two model examples (the solid line corresponds to spherical aberration correction while the dashed line corresponds to the chromatic aberration correction). The red curves show the ratio of the realistic probe size D$_{50}$ and the same quantity in the absence of the drill D$_{50 \mathrm{no-drill}}$ as a function of the drill diameter. For both model cases, we observe an increasing probe size (and decreasing ability to correct aberrations) with increasing drill diameter.}
    \label{fig:Fig_drill}
\end{figure}

The last element before the interaction is the mirror to couple the laser beam to the electron-optical axis. Since the mirror is planar, we simulate the laser axis as one straight line without bending. However, the mirror features a hole in the center necessary to let the electron beam pass through, which is represented only by a simple circular beam block in the simulations. The beam block has a diameter equal to the drill size and absorbs all intensity that hits it. The influence of these holes is analyzed in Fig.~\ref{fig:Fig_drill}. To not disturb the reflected laser and avoid losses of laser energy, it is beneficial to choose a small drill diameter below 1 mm; we considered drill diameters of 0.2 mm.

\section{Electron propagation}
\label{App:electron_propagation}
\subsection{Propagation through the optical field}
The propagation of electrons through the laser interaction volume is performed by numerically solving the relativistic equations of motion with the ponderomotive force [Eq.~\eqref{Eq:trajectory_eq}]. The problem is discretized in small time intervals $\Delta t$ and in each time interval, the electron is exposed to a certain gradient of laser intensity, which imposes a change in its momentum $\Delta\pb=\Delta t\,\Fb $, where $\Fb$ is the ponderomotive force. We further neglect the $z$ component of the gradient of the electric intensity and only evaluate the transverse ($x/y$) components of the ponderomotive force
\begin{align}
    F_{x/y}(\rb, t)
    &=-\frac{\alpha h \varepsilon_0 c}{2\gamma m_\mathrm{e} \omega_\mathrm{L}^2}\, \nabla_{x/y} \, \left( \lvert \Eb_\perp(\rb)\rvert^2 \right) \, g^2(t\pm (z-z_0)/c)
\end{align}
where $\alpha$ is the fine structure constant, $h$ is the Planck constant, $E_\mathrm{L}$ is the laser energy in a fixed $z$ plane, where the force is evaluated, $\omega_\mathrm{L}$ is the central laser angular frequency, $z_0$ is the $z$-coordinate of the central plane of the interaction volume, and $\gamma=1/\sqrt{1-v^2/c^2}$ is the Lorentz factor with $v$ the electron velocity. $\nabla_{x/y}$ denotes the partial derivative with respect to $x/y$ to calculate the corresponding component of the force. The $+/-$ sign in the denominator stands for the counter-/co-propagating electron and laser pulses. We then update the electron's position and continue in the iterative solution of the equation of motion until the electron enters the region where the light intensity approaches zero. 

We also tried to simplify the calculation by replacing many iterative steps through the laser-electron interaction volume with only one single-step interaction in the spirit of Ref.~\cite{Mihaila2022} and the paraxial approximation for both electron and laser pulses. This simplified approach still treats each electron individually. We find what is the laser beam distribution in a $xy$-plane along the optical axis, where an electron meets the peak power of a laser pulse. This laser distribution is then applied to the electron in one step with energy corresponding to the whole laser pulse energy, i.e., similarly to the action of electron lenses, we assume that the electron experiences a sudden momentum change and leaves the laser field with the velocity given by Eq.~\eqref{Eq:velocity_phase_lens}, where the corresponding phase is
\begin{align}
    \varphi (x,y)
 &=\frac{\pi \alpha \varepsilon_0 c}{\gamma m_\mathrm{e} \omega_\mathrm{L}^2(1\pm v/c)}
   \lvert \Eb_\perp(x,y)\rvert^2.
\end{align}

In the simulations, we further multiply the resulting velocity vector by a constant to ensure that the velocity magnitude is constant (due to the elastic nature of the interaction). When we compared the simplified approach and the iterative tracing approach, we found that both provide the same results, and the differences are negligible for our two examples. This is because the length of the interaction volume along the optical axis is much smaller than the distances between optical elements.

\subsection{Propagation through electron optics}
Within the thin-lens approximation, we assume that all electron lenses produce an abrupt change in electron momentum (velocity) in planes where the lenses are placed. By combining Eqs.~\eqref{Eq:velocity_phase_lens} and \eqref{eq:lens_focusing_spherical}, we find that for a non-aberrated lens (lens EL$_1$), the electron's velocity modified by the lens in the paraxial approximation becomes
\begin{align}
    \mathbf{v}\approx\left[\mathbf{v}_0-v_{z,0}\left(\frac{x}{f},\frac{y}{f},0\right)\right]\mathrm{const.},
    \label{Eq:lens_tracing}
\end{align}
while for an aberrated lens (lens EL$_2$) it reads
\begin{align}
    \mathbf{v}\approx\left[\mathbf{v}_0-v_{z,0}\left(\frac{x}{f(V)}+\frac{C_\mathrm{S}R^2x}{f(V)^4},\frac{y}{f(V)}+\frac{C_\mathrm{S}R^2y}{f(V)^4},0\right)\right]\mathrm{const.},
    \label{Eq:lens_tracing_aberrated}
\end{align}
where the constant is chosen such that the velocity magnitude is preserved, and the focal distance varies for electrons with energies different from the nominal one due to chromatic aberration according to Eq.~\eqref{eq:chromatic}.

\section{Laser setup optimization} 
\label{App:optimization}
The laser optimization procedure involves adjusting several parameters, including the SLM pattern, laser pulse energy, and electron objective lens defocus. Optimization is evaluated based on the D$_{50}$ metric related to the electron spot size. Before introducing any laser, the objective lens defocus is optimized to minimize D$_{50}$, with the corresponding result plotted as ``Aberrated No Correction'' in Fig.~\ref{fig:Fig2}(b) and Fig.~\ref{fig:Fig3}(b). This serves as the baseline reference to assess whether the laser can improve the D$_{50}$.

Subsequently, we apply a basic SLM pattern to generate a Laguere-Gaussian beam (LG$_{01}$ mode) and direct it into the microscope at low laser-pulse energies to monitor small variations in D$_{50}$. We then iteratively adjust the defocus and laser energy to achieve the lowest possible D$_{50}$. Once this is accomplished, the SLM pattern is further fine-tuned. Modifications to the SLM pattern are limited to changes in laser defocus and spherical aberration. This has the same effect as introducing a lens with a specific focal length and spherical aberration at the SLM plane. By adjusting the defocus, we shift the laser focal point along the $z$-axis within the interaction volume, enabling the electrons to interact with a diverging beam rather than a tightly focused one, as illustrated in Fig.~\ref{fig:Fig3}(a). The laser spherical aberration increases the laser depth of focus along the $z$-axis, as shown in Fig.~\ref{fig:Fig2}(a). These iterations were performed manually, which is inefficient, highlighting the need for advanced automation in future optimizations.

\section{Analytical solutions of ideal laser intensity distributions} 
\label{App:analytical_solutions}
Within the ray-tracing formalism, it is possible to find analytically the electrons' momenta in the plane just after the interaction region required for compensating for the aberrations of the lens EL$_2$. With such knowledge, we can deduce an approximate light intensity distribution in the interaction volume.

The electron momentum distribution can be obtained by back-tracing electrons from the desired perfect focal point at the position $R=0$ in the sample plane ($z=0$) through the lenses EL$_2$ and EL$_1$ to the plane $z=-(d_2+d_1+d_0)$ (see Fig.~\ref{fig:Fig1}). By assuming the lens parameters from Table \ref{tab:Tab1} set in such a way that the lens EL$_1$ images the plane $z=-(d_2+d_1+d_0)$ onto the plane $z=-d_2$ and thus $d_0=f_1 d_1/(d_1-f_1)$ holds, we find that the radial velocity component should ideally become
\begin{align}
    v_R\approx v_{z,0}\frac{Rd_1}{d_0}\left[\frac{d_1}{d_0}\left(-\frac{1}{d_2}+\frac{1}{f_2}+\frac{C_\mathrm{S}R^2d_1^2}{f_2^4 d_0^2}\right)-\frac{1}{f_1}\right].
\end{align}
We note that the terms $\propto R$ arise due to compensation of defocus while the terms $\propto R^3$ are to compensate the spherical aberration of the lens EL$_2$. To modify an initially collimated beam, which can be considered as a plane wave, into the required state we can calculate what will be a corresponding electron wave phase change $\varphi (R)$ caused by the laser:

\begin{align}
    \varphi(R) = -\frac{qR^2d_1}{2d_0} 
    \left[\frac{d_1}{d_0}\left(- \frac{1}{d_2} + \frac{1}{f_2}+ R^2
                    \frac{C_\mathrm{S}d_1^2 }{2 f_2^4 d_0^2}\right)
             - \frac{1}{f_1 }
           \right]  + \mathrm{const.}
    \label{eq:analytical_spherical}
\end{align}
The ideal phase (or based on Eq.~\eqref{Eq:velocity_phase_lens} the acquired radial velocity component) imprinted in a single interaction plane can be converted to the corresponding laser planar energy density
\begin{align}
    \eta(R) = \varphi(R)\, \frac{
        2 \pi\, (1 \pm v/c)\, E_\mathrm{e}
        }{
        \alpha\, \lambda_\mathrm{L}^2
        },
\label{eq:laser_planar_energy_density_spherical}
\end{align}
where $E_\mathrm{e}=\gamma m_\mathrm{e}c^2$ is the relativistic electron energy and $\lambda_\mathrm{L}$ is the central wavelength of the laser. The $+/-$ sign again corresponds to counter-/co-propagating laser and electron pulses. We note that although the ideal radial velocity is uniquely determined, it is not the case for the phase and magnitude of the electric field and the laser planar density as adding an arbitrary constant to the electric field still yields the same ponderomotive force. 

\begin{figure}
    \centering
    \includegraphics[width=0.5\linewidth]{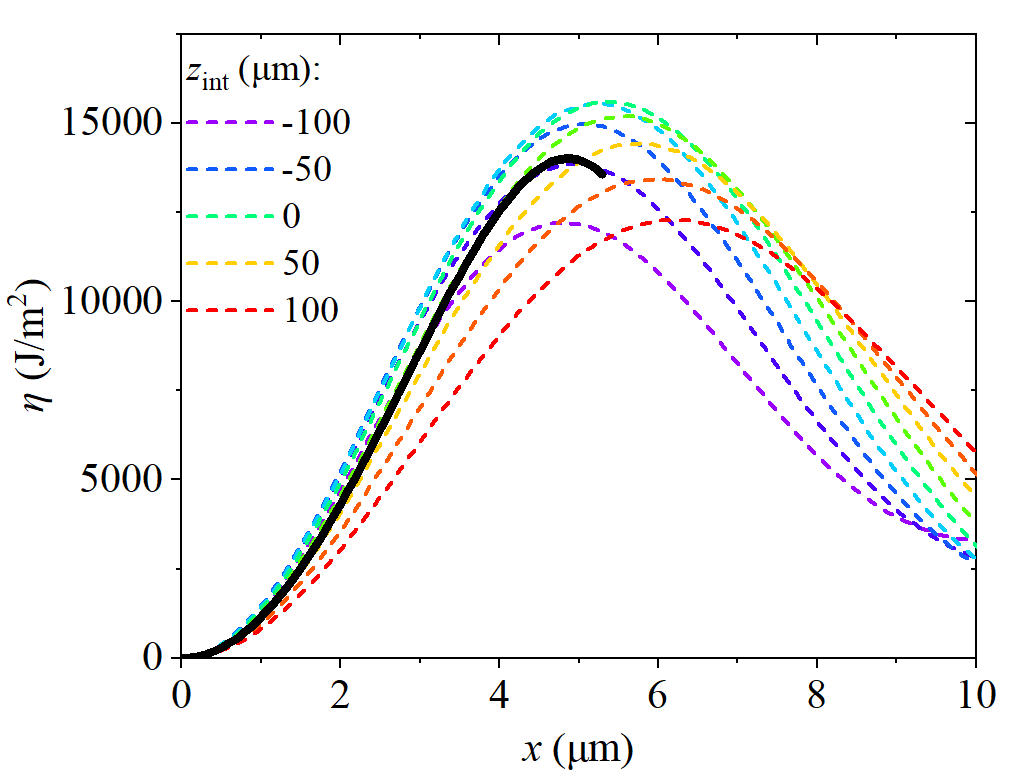}
    \caption{Laser planar energy density cross sections $\eta$ in various planes in interaction volume. Dashed lines correspond to cuts in selected planes from Fig.~\ref{fig:Fig2}(a). The black solid line shows the ideal laser profile calculated analytically by Eq.~\eqref{eq:laser_planar_energy_density_spherical}.}
    \label{fig:Fig_analytical_spherical}
\end{figure}

If we apply such an analytically calculated laser profile in Example 1 (spherical aberration correction), we get a probe diameter D$_{50}$ = 24 pm. This number is mainly determined by a chromatic aberration, which is not corrected in this case. However, in reality, the probe diameter will be enlarged by diffraction, which is not considered in our tracing approach. We also compare the idealized $\eta$ from Eq.~\eqref{eq:laser_planar_energy_density_spherical} with the numerically calculated counterparts in several planes in the interaction volume in  Fig.~\ref{fig:Fig_analytical_spherical}. 

\begin{figure}
    \centering
    \includegraphics[width=0.5\linewidth]{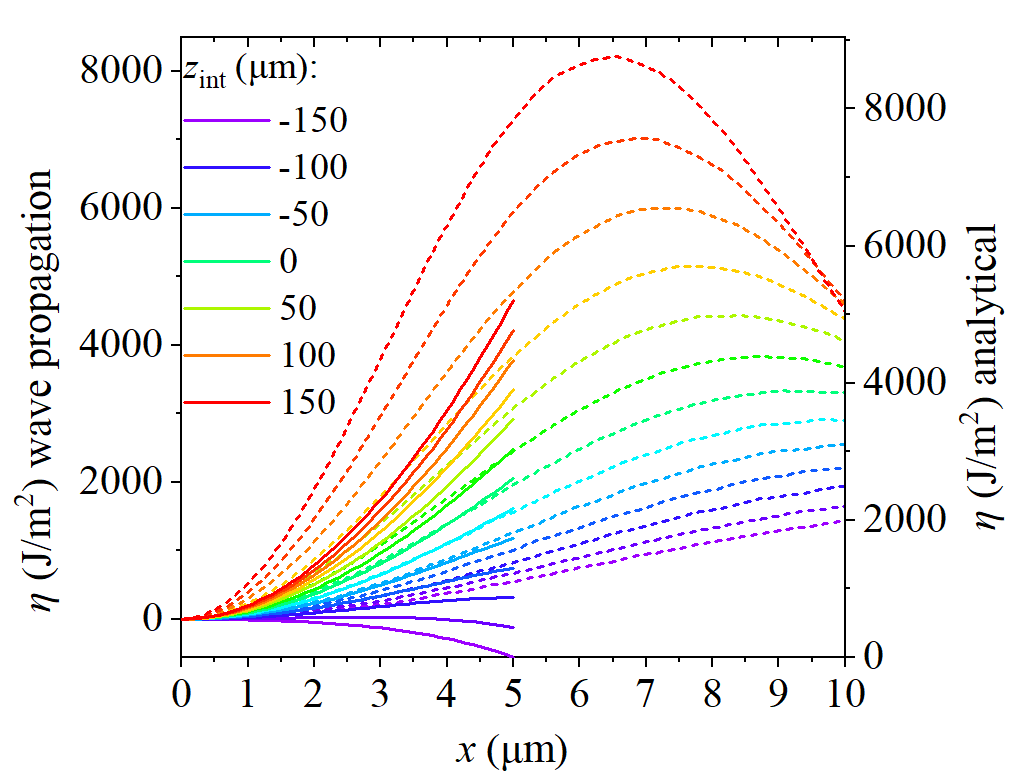}
    \caption{Laser planar energy density cross sections $\eta$ in various planes in interaction volume. Dashed lines correspond to cuts in selected planes from Fig.~\ref{fig:Fig3}(a). Solid lines show the ideal distribution of laser intensity calculated analytically by Eq.~\eqref{eq:laser_planar_energy_density_chromatic}.}
    \label{fig:Fig_analytical_chromatic}
\end{figure}

Similarly, we can analytically calculate the ideal laser-induced phase shift to correct chromatic aberration as shown in Fig.~\ref{fig:Fig_analytical_chromatic}. We can adapt Eq.~\eqref{eq:analytical_spherical} by considering that the electron wave vector and the focal distance $f_2$ depend on the electron energy. We can, in turn, find the effective distance along the $z_\mathrm{int}$ axis, where we need to apply the desired phase modulation. In reality, the distribution of electron energies along $z_\mathrm{int}$ is slightly mixed because electrons were emitted from the cathode during a nonzero time interval corresponding to the length of a photoemission laser pulse. However, we can suppose an ideal case for deriving an analytical solution of the best laser distribution: an infinitesimally short emission time interval. Then, the imprinted phase becomes
\begin{align}
    \varphi(R, z_\mathrm{int}) = &  -\frac{q(z_\mathrm{int})R^2d_1}{2d_0} 
        \left[\frac{d_1}{d_0}\left(- \frac{1}{d_2} + \frac{1}{f_2(z_\mathrm{int})}   +R^2 \frac{ C_\mathrm{S}d_1^2}{2 f_2^4(z_\mathrm{int}) \,d_0^2} \right)
            - \frac{1}{f_1}
         \right]
    + \mathrm{const},
\label{eq:analytical_chromatic}
\end{align}
where the focal distance is evaluated based on Eq.~\eqref{eq:chromatic}
\begin{align}
    f_2(z_\mathrm{int})=  f_0 + C_\mathrm{C} \frac{V(z_\mathrm{int}) - V_0}{V_0},
\end{align}
and where
\begin{align}
   V(z_\mathrm{int})  = \frac{m_\mathrm{e} \, c^2}{e}  \left( -1 + \sqrt{\frac{1}{
        1-  \left(
                \frac{z_\mathrm{int} + d_\mathrm{ini}}{d_\mathrm{ini}}
            \right)^2
            \left(
                1-
                \left(
                1+ \frac{e \, V_0}{m_\mathrm{e} c^2}
                \right)^{-2}
            \right)
    }}  \right),
\label{eq:analytical_acc_rel}
\end{align}
which can be simplified if we omit the relativistic correction to $V(z_\mathrm{int}) = V_0 ((z_\mathrm{int} + d_\mathrm{ini})/d_\mathrm{ini})^2$. In Eq.~\eqref{eq:analytical_acc_rel}, we observe that the dependence on electron energy spread is absent. That happens because higher-energy electrons propagate faster, and thus are expected to be located further along the $z_\mathrm{int}$-axis at the time of interaction. In contrast, lower-energy electrons propagate slower and are expected to be positioned closer to the source within the interaction volume.  For instance, an electron with nominal acceleration energy $V = V_0$ will interact with the peak of the laser intensity envelope at $z_\mathrm{int} = 0$. Eq.~\eqref{eq:analytical_acc_rel} can also be used to evaluate the electron wave vector
\begin{align}
q(z_\mathrm{int}) = \frac{
        2 \pi
        \sqrt{
            2 m_\mathrm{e} e V(z_\mathrm{int}) \left(
                1+\frac{e}{2 m_\mathrm{e} c^2} V(z_\mathrm{int})
            \right)
        }
    }
    {h}.
\end{align}

The corresponding laser planar energy density in Example 2 (chromatic aberration correction) then becomes 
\begin{align}
    \eta(R, z_\mathrm{int}) = \varphi(R, z_\mathrm{int})\, \frac{
        2 \pi\, (1 \pm v/c)\, E_\mathrm{e}(z_\mathrm{int})
        }{
        \alpha\, \lambda_\mathrm{L}^2
        }.
\label{eq:laser_planar_energy_density_chromatic}
\end{align}

If we apply the analytically calculated laser profile in Example 2 (chromatic aberration correction) from Eq.~\eqref{eq:laser_planar_energy_density_chromatic}, we get the electron probe diameter D$_{50}$ = 0.4 nm. However, in reality, the diameter will be increased by diffraction which is not considered in our model.

\end{document}